\documentclass[a4paper]{panl}
\usepackage{cite}
\usepackage{wrapfig}
\usepackage{graphicx}
\usepackage{amssymb}
\usepackage{amsfonts}
\usepackage{amsmath}
\usepackage{longtable}
\usepackage{rotating}
\usepackage{lscape}
\usepackage{epsfig}
\usepackage{multirow}

\usepackage{cmap}
\usepackage{epstopdf}
\usepackage{braket}
\usepackage{bm}
\usepackage[percent]{overpic}
\usepackage{mathrsfs}
\usepackage{amsmath}
\usepackage{graphicx}
\usepackage{dcolumn}
\usepackage{latexsym}
\usepackage{amssymb}
\usepackage{eqnarray} 
\usepackage{tabularx}
\usepackage{capt-of}
\usepackage{csquotes}
\usepackage{ragged2e} 
\usepackage{subfigure}

\originalTeX
\begin{document}

\title{Field-free molecular alignment by the optimized two-color laser fields}
\maketitle
\authors{E.\,A.\,Koval$^{a}$\footnote{E-mail: e-cov@yandex.ru}}
\setcounter{footnote}{0}
\from{$^{a}$\,Bogoliubov Laboratory of Theoretical Physics, Joint Institute for Nuclear Research, Dubna, Moscow Region 141980, Russian Federation}

\begin{abstract}

We have theoretically investigated the molecular orientation by a asymmetric potential created by the superposition of two-color laser fields. The time-dependent Schrodinger equation is solved numerically for different field parameters. We have shown how enhancement or suppression of the molecular orientation can be manipulated by the laser field parameters, such as time between laser pulses, the different intensity of the pulses, etc. 
\end{abstract}
\vspace*{6pt}


\label{SecIntro}
\section*{Introduction}

Recently, a rapid growth of theoretical \cite{mellado2020linear, wang2020optimal, mun2019orientation, kanai2001numerical} and experimental \cite{mun2019orientation, zhang2011field, ohmura2004quantum, oda2010all} studies of the alignment in laser and combined fields have been seen~\cite{koch2019quantum}. Molecular alignment control methods, for example, are practically used in applications such as multiphoton ionization\cite{suzuki2004optimal}, angular distribution of photoelectrons~\cite{holmegaard2010photoelectron}, high harmonics generation~\cite{itatani2004tomographic}.

In our article we investigate field-free molecular alignment by a two-color laser field. a good agreement with the results of other authors were obtained and the impact of time delay between the pulses and the different ratios of the pulses' amplitudes on the alignment cosine were investigated.
These results are important to the operations on a quantum qubit, based on the polar molecules in the optical lattices.

\label{SecMain}
\section*{Theory}

The Hamiltonian of the linear molecule in external field reads the form
\begin{equation}\label{H_genform}
H' = B \mathbf{J}^2 + V'(\theta,t) 
\end{equation}
within a rigid rotor approximation. 
$B = \hbar^2/2I$  is the rotational constant; $I$ denotes the moment of inertia; 
\begin{equation}\label{angular_mom}
\mathbf{J}^2 = -\frac{1}{\sin\theta} \frac{\partial}{\partial \theta} \left(\sin\theta \frac{\partial}{\partial\theta} \right) - \frac{1}{\sin^2 \theta} \frac{\partial^2}{\partial\phi^2} 
\end{equation}
is the operator of the angular momentum squared. $\theta \in [0,\pi]$ and $\phi \in [0,2\pi]$ are the polar and azimuthal angles~\cite{Arfken2005}. 

For a dimensionless Hamiltonian 
\begin{equation}\label{Hdimensionless}
H \equiv H' / B =\mathbf{J}^2 + V(\theta,t)
\end{equation}
and dimensionless time $t \equiv  Bt'/\hbar $ 
,  the corresponding time-dependent Schrodinger equation (TDSE) takes the form
\begin{equation}\label{scheqn}
i \frac{\partial}{\partial t} \psi(\theta,\phi,t) = \left( -\frac{1}{\sin\theta} \frac{\partial}{\partial \theta} \left(\sin\theta \frac{\partial}{\partial\theta} \right) - \frac{1}{\sin^2 \theta} \frac{\partial^2}{\partial\phi^2} 
+  V(\theta,t) \right) \psi(\theta,\phi,t)
\end{equation}

The interaction potential $V \equiv V'/B$ has the following form:
\begin{align}\label{Vgenform}
V(\theta,\tau) = -\mu\cos\theta\,F(t) - \frac1{2}\left(\Delta\alpha\cos^2\theta+\alpha_\bot\right)F^2(t)
\end{align}
where $\theta$ is the Euler angle between the internuclear molecular axis and the $Z$-axis in the laboratory frame; $\mu$  the permanent electric dipole moment; $\Delta\alpha=\alpha_\parallel-\alpha_\bot$ 
the polarizability anisotropy, with  $\alpha_\bot$ and $\alpha_\parallel$ being its perpendicular 
and parallel components.

One or two-color laser field linearly polarized along the laboratory fixed frame $Z$-axis is considered. 
The corresponding electric field ${\bf F}(t)=F(t){\bf Z}$ is given by the biharmonic function
\begin{equation}
F(t)=F_1(t)\cos(\omega (t-t_1))+F_{2}(t)\cos(2\omega (t-t_2)), 
\label{ElectricField}
\end{equation}
with $\omega$ and $2\omega$, $F_1(t)$ and $F_{2}(t)$, $t_1$ and $t_2$ being  the laser frequency, electric field strength (laser pulse envelope) and laser pulse time delay of corresponding harmonic. Hence, the solution of the TDSE (\ref{scheqn}) depends on the dimensionless interaction parameter $\Delta\omega=\frac{\max(F_1(t))^2\Delta\alpha}{2B}$. 
These parameters allow for comparing the dynamics that results from kicks of different shapes, lengths, and strengths.

Note that the energies are expressed in terms of the rotational constant $B$ and that a rotational period amounts to $\tau_r=\pi$.

The time-dependent Schr\"odinger equation is solved by the combining the Strang-Marchuk split-operator method~\cite{marchuk1990splitting} for the time variable, and the Discrete Variable Representation approach to the solving the Schr\"odinger equation on each time step.

We assume that at $t=0$ the molecule is in a field-free eigenstate $\psi(\Omega,t=0)=Y_{J,M}(\Omega)$, which is the initial step for the numerical solution of the time-dependent Schr\"odinger equation. the gaussian shape of the laser field envelopes is used:
\begin{equation}
F_i(t)=\exp[-2\ln2\frac{t^{2}}{{\tau_{FWHM,i}}^{2}}], i={1,2}.
\end{equation} 

In this work, the field-dressed rotational dynamics is analyzed in terms of the alignment expectation values (alignment cosine) 
\begin{equation}
\langle\cos^2\theta \rangle=\int\psi^*(\Omega,t)\cos^2\theta\psi(\Omega,t)d\Omega,
\label{eq:expdcg}
\end{equation}
with $\psi(\Omega,t)$ being the time-dependent wave function.
The wave function  $\psi(\Omega,t)$  and the expectation values depend on the 
laser field parameters~\eqref{ElectricField}. 

\label{SecResults}
\section*{Results}
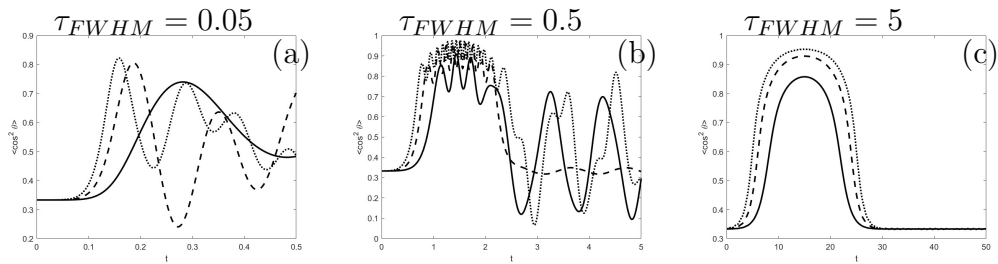
\begin{figure}[h!]
\begin{minipage}[t][0.2\textheight][t]{\textwidth}
		\begin{overpic}[height=3.3cm]{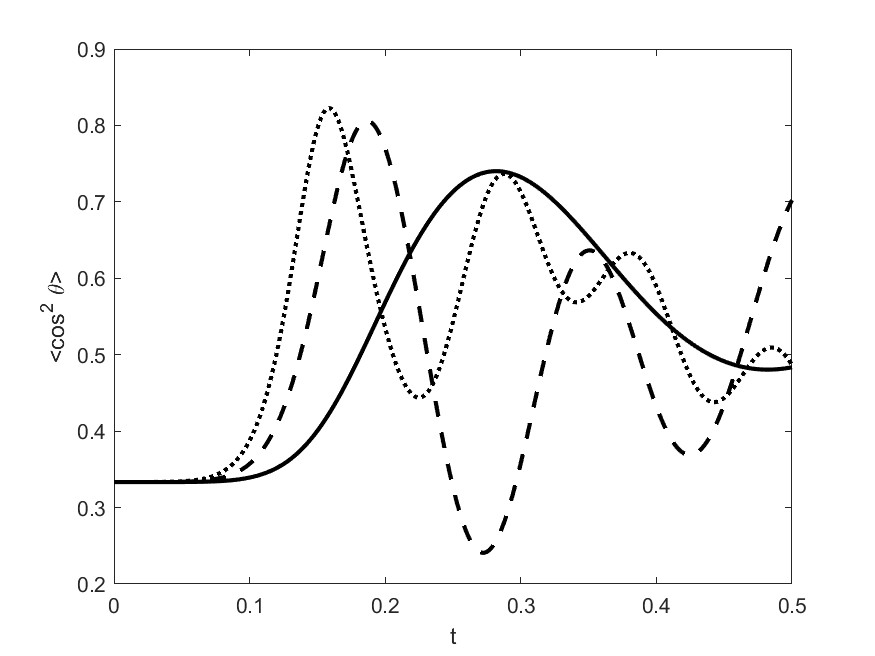}
                    \put(15,71){
                    {{$\tau_{FWHM}=0.05$}} 
                    }
                    \put(80,61){
                    {{(a)}} 
                    }
                \end{overpic}
		\begin{overpic}[height=3.3cm]{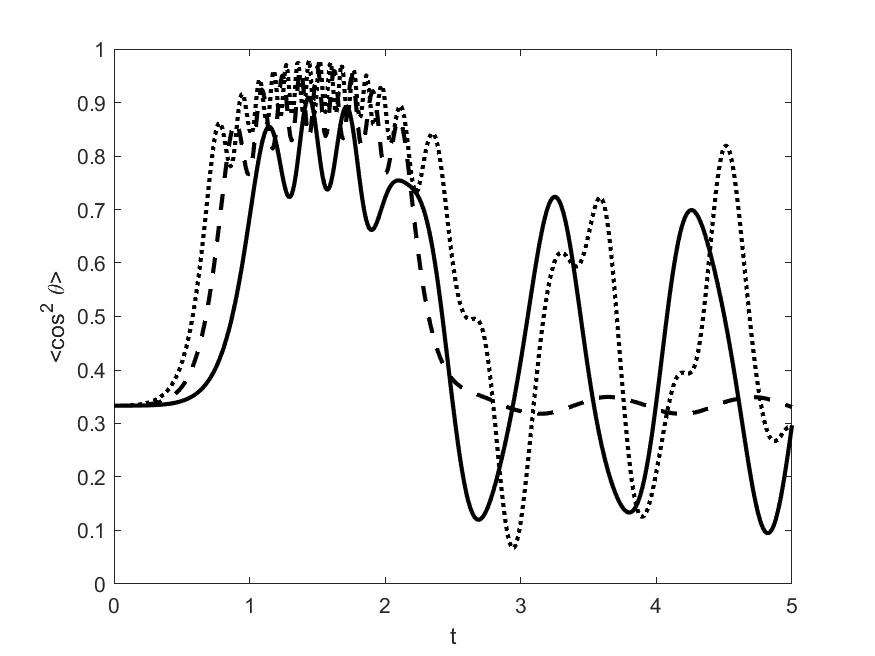}
                    \put(15,71){
                    {{$\tau_{FWHM}=0.5$}} 
                    }
                    \put(80,61){
                    {{(b)}} 
                    }
                \end{overpic}
		\begin{overpic}[height=3.3cm]{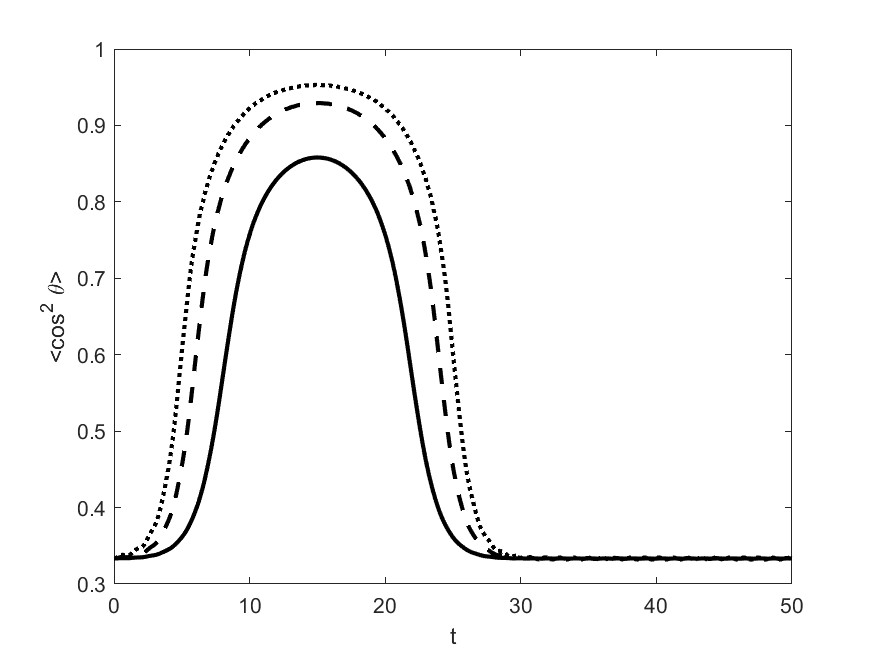}
                    \put(15,71){
                    {{$\tau_{FWHM}=5$}} 
                    }
                    \put(80,61){
                    {{(c)}} 
                    }
                \end{overpic}
\end{minipage}	
\vspace*{-1.5cm}
\captionof{figure}{
The alignment expectation value $\langle\cos^2(\theta)\rangle$  for the one-color pulse for different pulse duration. $\Delta\omega=100$ is marked by the solid line, $\Delta\omega=400$ --- by the dashed line, $\Delta\omega=900$ is marked by the dot-dashed line. The pulse durations are as follows: $\tau_{FWHM}=0.05$(a), $\tau_{FWHM}=0.5$(b), $\tau_{FWHM}=5$(c).
}
\label{figLikeOrtigoso}
\end{figure}

In order to verify the computational scheme we have reproduced the results for the one-color gaussian pulse with different pulse durations and model molecule parameters of the Ref.\cite{ortigoso1999time}, presented in the Fig.~\ref{figLikeOrtigoso}. The good agreement with results of the Ref.\cite{ortigoso1999time} was obtained for molecules alignment with one-color pulse. The influence of pulses with longer duration show almost adiabatic response of the rigid rotor. While shorter pulses lead to the strongly nonadiabatic behaviour with oscillating alignment even after pulse switch-off. The stronger laser field apmlitudes allow to get higher mplecules alignment during the pulse. The nonadiabatic behaviour also shows revivals of the alignment after short pulses.

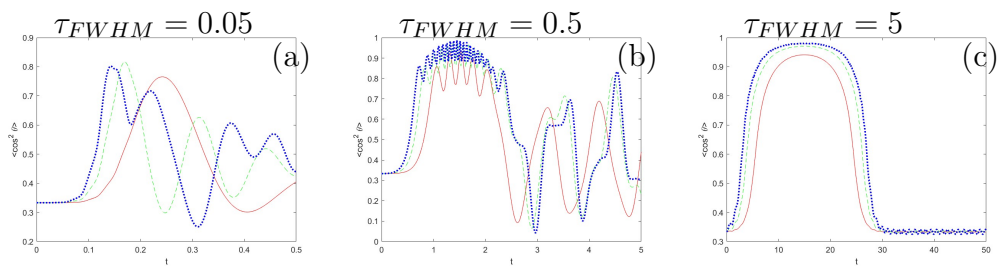
\begin{figure}[h!]
\begin{minipage}[t][0.2\textheight][t]{\textwidth}
		\begin{overpic}[height=3.3cm]{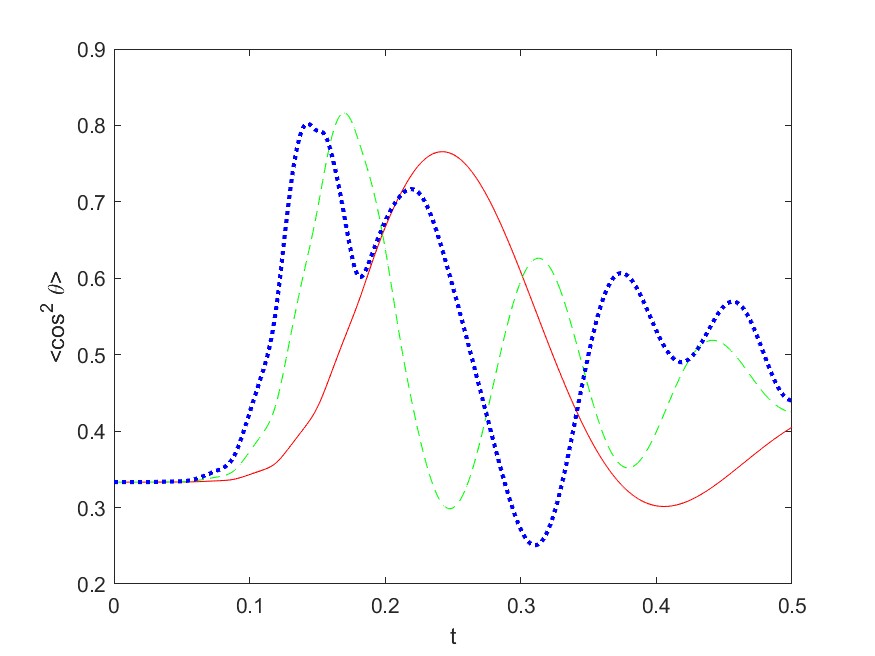}
                    \put(15,71){
                    {{$\tau_{FWHM}=0.05$}} 
                    }
                    \put(80,61){
                    {{(a)}} 
                    }
                \end{overpic}
		\begin{overpic}[height=3.3cm]{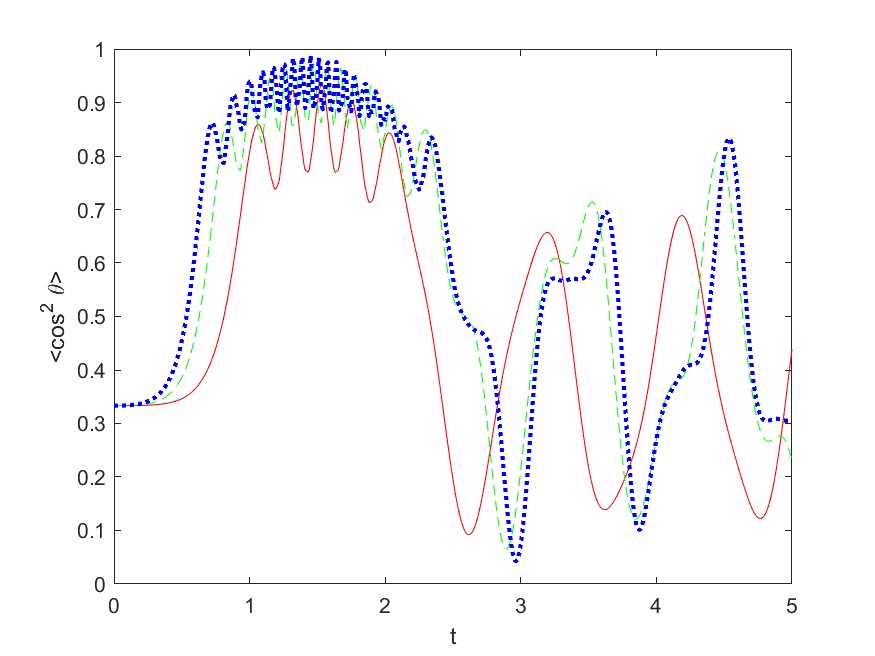}
                    \put(15,71){
                    {{$\tau_{FWHM}=0.5$}} 
                    }
                    \put(80,61){
                    {{(b)}} 
                    }
                \end{overpic}
		\begin{overpic}[height=3.3cm]{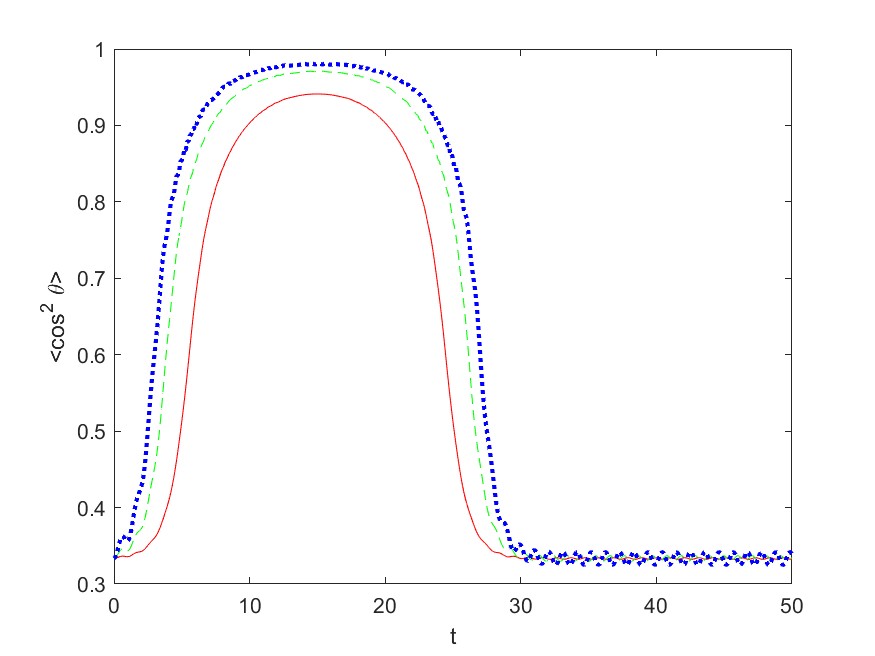}
                    \put(15,71){
                    {{$\tau_{FWHM}=5$}} 
                    }
                    \put(80,61){
                    {{(c)}} 
                    }
                \end{overpic}
\end{minipage}	
\vspace*{-1.5cm}
\captionof{figure}{
The alignment cosine $\langle\cos^2(\theta)\rangle$  for the {\bf two-color} pulse for different pulse duration. $\Delta\omega=100$ is marked by the solid line, $\Delta\omega=400$ --- by the dashed line, $\Delta\omega=900$ is marked by the dot-dashed line. The pulse durations are as follows: $\tau_{FWHM}=0.05$(a), $\tau_{FWHM}=0.5$(b), $\tau_{FWHM}=5$(c).
}
\label{figTwoColorDifferentPulseDuration}
\end{figure}

For two-color parallel laser fields the number of alignment cosine oscillations is increased comparing to the the one-color case, as shown in Fig.\ref{figTwoColorDifferentPulseDuration}. While shapes of the dependences of the alignment cosine on the pulses duration are close to the one-color case, the period of the alignment are longer for the same pulse duration as in one-color case. 
      
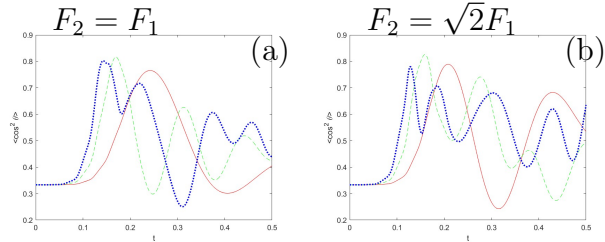
\begin{figure}[h!]
\begin{minipage}[t][0.2\textheight][t]{\textwidth}
\centering
		\begin{overpic}[height=3.cm]{imgRes/alignment_TwoColorPulse_FWHM=0.05.jpg}
                    \put(15,71){
                    {{$F_{2}=F_1$}} 
                    }
                    \put(80,61){
                    {{(a)}} 
                    }
                \end{overpic}
		 \begin{overpic}[height=3.cm]{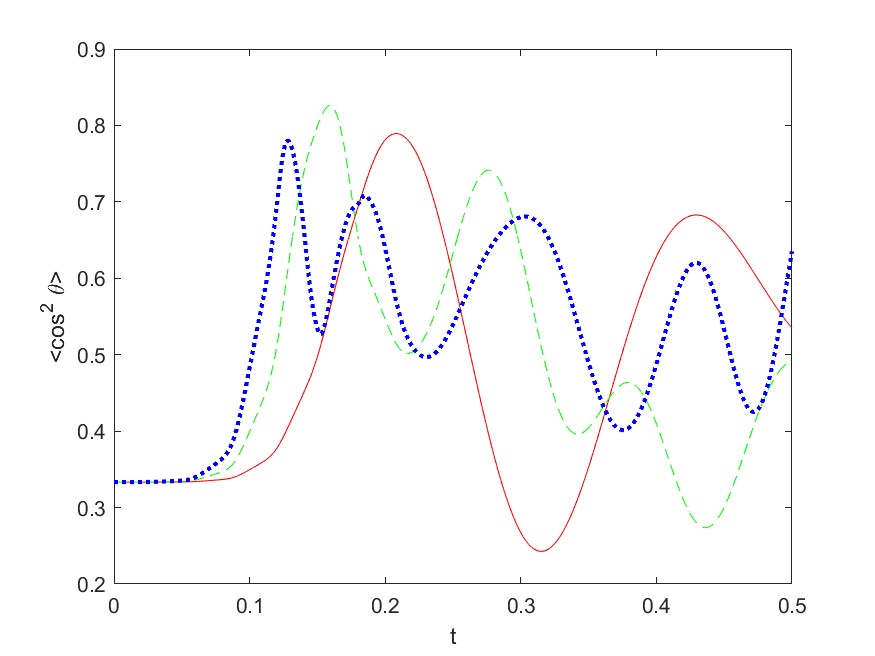}
                    \put(15,71){
                    {{$F_{2}=\sqrt{2}F_1$}} 
                    }
                    \put(80,61){
                    {{(b)}} 
                    }
                \end{overpic}
\end{minipage}	
\vspace*{-1.5cm}
\captionof{figure}{
The alignment cosine $\langle\cos^2(\theta)\rangle$  for the {\bf two-color} pulse for different pulse intensities' ratio. $\Delta\omega=100$ is marked by the solid line, $\Delta\omega=400$ --- by the dashed line, $\Delta\omega=900$ is marked by the dot-dashed line. The pulse intensities' ratio are: $F_{2}=F_1$(a), $F_{2}=\sqrt{2}F_1$(b). The FWHM of the pulses is fixed: $\tau_{FWHM}=0.05$.
}
\label{figTwoColorDifferentAmplitudeRatio}
\end{figure}

We also calculated the results for the case when the intensity ($I=|F^2|$) of the second harmonic pulse is two times higher than first harmonic $I_2 = 2 I_1$. Fig.\ref{figTwoColorDifferentAmplitudeRatio} illustrates obtained results and shows that increasing of the intensity of the second harmonic pulse leads to the longer standing molecular alignment for the short pulse durations.

\begin{figure}[h!]
\begin{minipage}[t][0.2\textheight][t]{\textwidth}
		\begin{overpic}[height=3.3cm]{imgRes/alignment_TwoColorPulse_FWHM=0.05.jpg}
                    \put(15,71){
                    {{$t_2=t_1$}} 
                    }
                    \put(80,61){
                    {{(a)}} 
                    }
                \end{overpic}
		  \begin{overpic}[height=3.3cm]{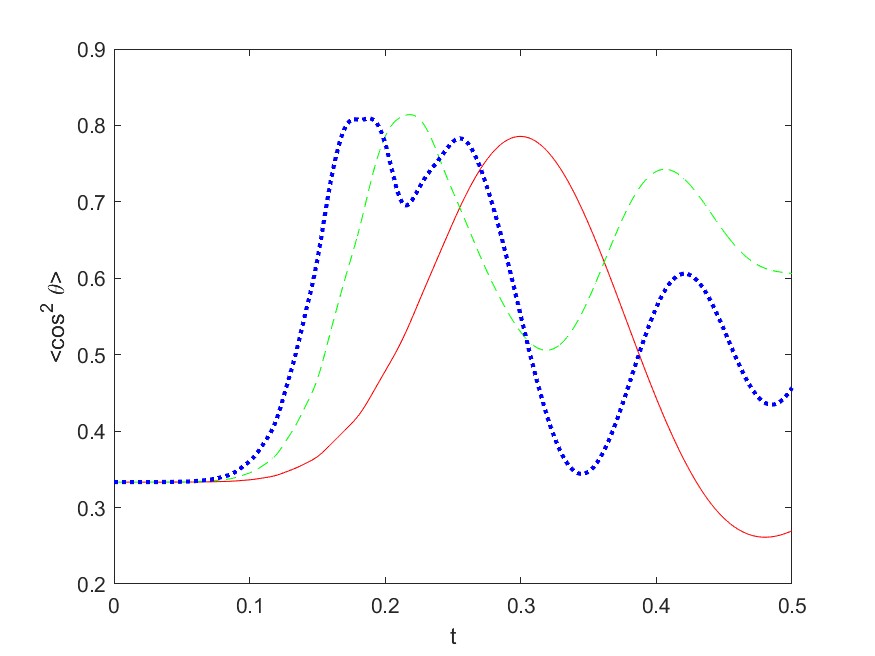}
                    \put(15,71){
                    {{$t_2=1.5t_1$}} 
                    }
                    \put(80,61){
                    {{(b)}} 
                    }
                \end{overpic}
                \begin{overpic}[height=3.3cm]{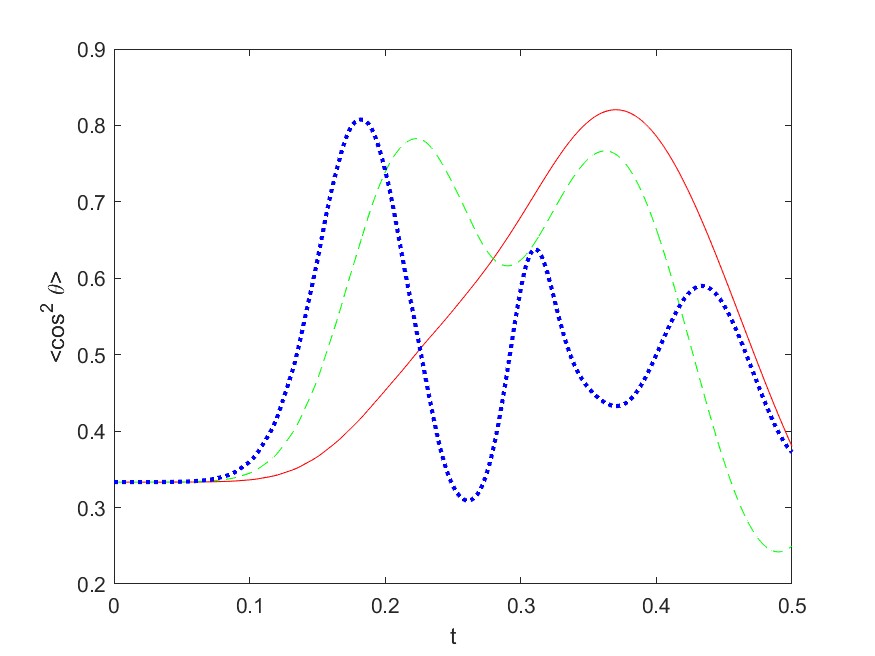}
                    \put(15,71){
                    {{$t_2=2t_1$}} 
                    }
                    \put(80,61){
                    {{(c)}} 
                    }
                \end{overpic}
\end{minipage}	
\vspace*{-1.5cm}
\captionof{figure}{
The alignment cosine $\langle\cos^2(\theta)\rangle$  for the {\bf two-color} pulse for different second harmonic pulse delays. $\Delta\omega=100$ is marked by the solid line, $\Delta\omega=400$ --- by the dashed line, $\Delta\omega=900$ is marked by the dot-dashed line. The second harmonic pulse time delays are : $t_2=t_1$(a), $t_2=1.5t_1$(b), $t_2=2t_1$(c). The FWHM of the first harmonic pulse is fixed: $\tau_{FWHM}=0.05$.
}
\label{figTwoColorDifferentTimeDelays}
\end{figure}
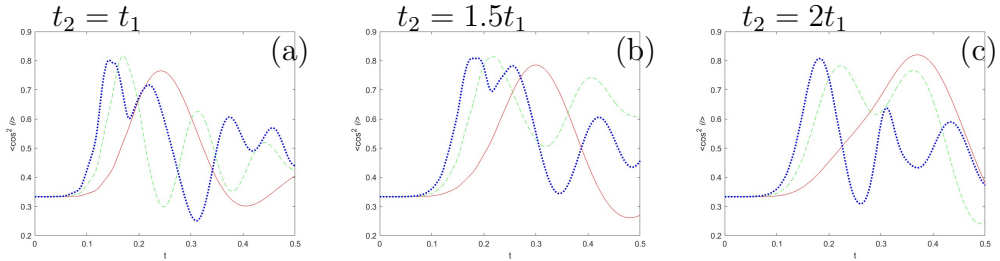

The variation of the second harmonic pulse delay (relative to the first harmonic pulse), as shown in the Fig.\ref{figTwoColorDifferentTimeDelays}, allows us to change interfering pattern of the pulses on the wave function, in order to increase the duration of the alignment for short pulses.

\bibliographystyle{pepan}
\bibliography{Koval_biblio} 

\begin{thebibliography}{10}
\def\selectlanguageifdefined#1{
\expandafter\ifx\csname date#1\endcsname\relax
\else\selectlanguage{#1}\fi}
\providecommand*{\href}[2]{{\small #2}}
\providecommand*{\url}[1]{{\small #1}}
\providecommand*{\BibUrl}[1]{\url{#1}}
\providecommand{\BibAnnote}[1]{}
\providecommand*{\BibEmph}[1]{\emph{#1}}
\ProvideTextCommandDefault{\cyrdash}{\hbox to.8em{--\hss--}}
\providecommand*{\BibDash}{\ifdim\lastskip>0pt\unskip\nobreak\hskip.2em\fi
\cyrdash\hskip.2em\ignorespaces}

\bibitem{mellado2020linear}
\selectlanguageifdefined{english}
\BibEmph{Mellado-Alcedo D., Quintero N.R., Gonz{\'a}lez-F{\'e}rez R.} Linear
  polar molecule in a two-color cw laser field: A symmetry analysis~// Physical
  Review A. \BibDash
\newblock 2020. \BibDash
\newblock V. 102, no.~2. \BibDash
\newblock P.~023110.

\bibitem{wang2020optimal}
\selectlanguageifdefined{english}
\BibEmph{Wang S., Henriksen N.E.} Optimal field-free molecular orientation with
  nonresonant two-color adiabatic-turn-on and sudden-turn-off laser pulses~//
  Physical Review A. \BibDash
\newblock 2020. \BibDash
\newblock V. 102, no.~6. \BibDash
\newblock P.~063120.

\bibitem{mun2019orientation}
\selectlanguageifdefined{english}
\BibEmph{Mun J.H., Sakai H., Gonz{\'a}lez-F{\'e}rez R.} Orientation of linear
  molecules in two-color laser fields with perpendicularly crossed
  polarizations~// Physical Review A. \BibDash
\newblock 2019. \BibDash
\newblock V.~99, no.~5. \BibDash
\newblock P.~053424.

\bibitem{kanai2001numerical}
\selectlanguageifdefined{english}
\BibEmph{Kanai T., Sakai H.} Numerical simulations of molecular orientation
  using strong, nonresonant, two-color laser fields~// The Journal of Chemical
  Physics. \BibDash
\newblock 2001. \BibDash
\newblock V. 115, no.~12. \BibDash
\newblock P.~5492--5497.

\bibitem{zhang2011field}
\selectlanguageifdefined{english}
\BibEmph{Zhang S., Lu C., Jia T., Wang Z., Sun Z.} Field-free molecular
  orientation enhanced by two dual-color laser subpulses~// The Journal of
  chemical physics. \BibDash
\newblock 2011. \BibDash
\newblock V. 135, no.~3. \BibDash
\newblock P.~034301.

\bibitem{ohmura2004quantum}
\selectlanguageifdefined{english}
\BibEmph{Ohmura H., Nakanaga T.} Quantum control of molecular orientation by
  two-color laser fields~// The Journal of chemical physics. \BibDash
\newblock 2004. \BibDash
\newblock V. 120, no.~11. \BibDash
\newblock P.~5176--5180.

\bibitem{oda2010all}
\selectlanguageifdefined{english}
\BibEmph{Oda K., Hita M., Minemoto S., Sakai H.} All-optical molecular
  orientation~// Physical review letters. \BibDash
\newblock 2010. \BibDash
\newblock V. 104, no.~21. \BibDash
\newblock P.~213901.

\bibitem{koch2019quantum}
\selectlanguageifdefined{english}
\BibEmph{Koch C.P., Lemeshko M., Sugny D.} Quantum control of molecular
  rotation~// Reviews of Modern Physics. \BibDash
\newblock 2019. \BibDash
\newblock V.~91, no.~3. \BibDash
\newblock P.~035005.

\bibitem{suzuki2004optimal}
\selectlanguageifdefined{english}
\BibEmph{Suzuki T., Minemoto S., Kanai T., Sakai H.} Optimal control of
  multiphoton ionization processes in aligned I 2 molecules with time-dependent
  polarization pulses~// Physical review letters. \BibDash
\newblock 2004. \BibDash
\newblock V.~92, no.~13. \BibDash
\newblock P.~133005.

\bibitem{holmegaard2010photoelectron}
\selectlanguageifdefined{english}
\BibEmph{Holmegaard L., Hansen J.L., Kalh{\o}j L., Louise~Kragh S., Stapelfeldt
  H., Filsinger F., K{\"u}pper J., Meijer G., Dimitrovski D., Abu-Samha M.,
  others.} Photoelectron angular distributions from strong-field ionization of
  oriented molecules~// Nature Physics. \BibDash
\newblock 2010. \BibDash
\newblock V.~6, no.~6. \BibDash
\newblock P.~428--432.

\bibitem{itatani2004tomographic}
\selectlanguageifdefined{english}
\BibEmph{Itatani J., Levesque J., Zeidler D., Niikura H., P{\'e}pin H., Kieffer
  J.C., Corkum P.B., Villeneuve D.M.} Tomographic imaging of molecular
  orbitals~// Nature. \BibDash
\newblock 2004. \BibDash
\newblock V. 432, no. 7019. \BibDash
\newblock P.~867--871.

\bibitem{Arfken2005}
\selectlanguageifdefined{english}
\BibEmph{Arfken G., Weber H.} Mathematical Methods For Physicists International
  Student Edition. \BibDash
\newblock Elsevier Science, 2005.

\bibitem{marchuk1990splitting}
\selectlanguageifdefined{english}
\BibEmph{Marchuk G.I.} Splitting and alternating direction methods~// Handbook
  of numerical analysis. \BibDash
\newblock 1990. \BibDash
\newblock V.~1. \BibDash
\newblock P.~197--462.

\bibitem{ortigoso1999time}
\selectlanguageifdefined{english}
\BibEmph{Ortigoso J., Rodr{\i}guez M., Gupta M., Friedrich B.} Time evolution
  of pendular states created by the interaction of molecular polarizability
  with a pulsed nonresonant laser field~// The Journal of chemical physics.
  \BibDash
\newblock 1999. \BibDash
\newblock V. 110, no.~8. \BibDash
\newblock P.~3870--3875.

\end{thebibliography}

\end{document}